\documentclass[aps,prd,10pt,twocolumn,nofootinbib]{revtex4}

\usepackage{epsfig}
\usepackage{bm}
\usepackage{latexsym}
\usepackage{natbib}
\usepackage{url}
\usepackage{dcolumn}
\usepackage{color}
\usepackage{amsfonts,amssymb,amsmath}
\usepackage{graphicx,epsfig}
\usepackage{psfrag}
\usepackage{subfigure}
\usepackage{hyperref}
\hypersetup{colorlinks=true}
\usepackage{mathtools}
\usepackage{enumitem}
\usepackage{float}
\usepackage[dvipsnames]{xcolor}

\def\be{\begin{equation}}
\def\ee{\end{equation}}
\def\beq{\begin{equation}}
\def\eeq{\end{equation}}
\def\bea{\begin{eqnarray}}
\def\eea{\end{eqnarray}}
\def\LCDM{$\Lambda$CDM}

\begin{document}

\title{$H_0$ as a Universal FLRW Diagnostic}

\author{
Chethan Krishnan$^{a,}$\footnote{chethan.krishnan@gmail.com}, and Ranjini Mondol$^{a,}$\footnote{ranjinim@iisc.ac.in}
}

\affiliation{$^a$ Centre for High Energy Physics, Indian Institute of Science, Bangalore 560012, India
}

\begin{abstract}
We reverse the logic behind the apparent existence of $H_0$-tension, to design diagnostics for cosmological models. The basic idea is that the non-constancy of $H_0$ inferred from observations at different redshifts is a null hypothesis test for models within the FLRW paradigm -- if $H_0$ runs, the model is wrong. Depending on the kind of observational data, the most suitable form of the diagnostic can vary. As examples, we present two $H_0$ diagnostics that are adapted to two different BAO observables. We use these and the corresponding BAO data to Gaussian reconstruct the running of $H_0$ in flat \LCDM\ with Planck values for the model parameters. For flat \LCDM\ when the radiation contribution can be neglected, with comoving distance data, the diagnostic is a simple hypergeometric function. Possible late time deviations from the FLRW paradigm can also be accommodated, by simply keeping track of the (potentially anisotropic) sky variation of the diagnostic.

\end{abstract}

\maketitle

\section{Introduction}

Observations of the Hubble parameter $H(z)$ at various redshifts,  together with the matter content of the flat $\Lambda$CDM model as a function of $z$, yield a prediction for the present day Hubble parameter $H_0$ via the Friedmann equation: 
\bea H_0^2=\frac{H^2(z)}{\Omega_{m0}(1+z)^3+1-\Omega_{m0}}. \label{start} 
\eea
In $\Lambda$CDM (and indeed within any single FLRW model), $H_0$ is a constant by definition: it is the Hubble scale of the present epoch. But when one computes $H_0$ in the above manner from observations of $H(z)$ at $z \sim 1100$, it does not match with what is found from direct observations at $z\approx 0$ \cite{Intro0,Intro1,Intro2,Intro3,Intro4,Intro5,Intro6,Intro7,Intro8, Intro9,Knox:2019rjx,Vagnozzi}. This apparent red-shift dependence of a quantity that should tautologically be a constant, is called {\em $H_0$-tension}. 

Assuming that our working hypothesis of an FLRW framework is correct, the reason for the existence of the tension in \eqref{start} is of course, simple. The non-constancy of the right hand side of Eq. \eqref{start}  is a result of the mismatch between observations, and a model that is supposed to fit those observations. The observations are  captured by the numerator $H(z)$, and the model is defined by the $\Omega$'s in the denominator. When viewed this way, \eqref{start} is in fact a {\em definition} of $H_0$ as inferred from data at redshift $z$ and therefore it need not identically be a constant. The crucial point here is that there is no direct connection that data at finite redshift has, to zero redshift, other than through a model\footnote{A minor exception to this exists at very low red-shifts where one can use a Taylor expansion for the scale factor, without relying on explicit models. This is sometimes called cosmography, and is model independent within the FLRW framework. But if we want to compare results from sufficiently different redshifts, one necessarily needs a model.}.


Viewed this way as a tension between observations and models, there is no inherent contradiction in the non-constancy of the right hand side of Eq. \eqref{start} -- observations can be erroneous, and models can be wrong. For the purposes of this paper we will adopt the standpoint that observations are correct, and will use the potential non-constancy of $H_0$, as a diagnostic for the validity of models.

A point worthy of note regarding this perspective is that it begs us to consider the possibility that the right hand side of Eq. \eqref{start} is a (quasi-)continuous\footnote{The prefix ``quasi-" is supposed to capture the fact that observations are usually finite in number. The point is that we are limited only by the observational resolution in redshift.} function of $z$. After all, if the mismatch between model and observation is a real effect, then it stands to reason that the value of $H(z)$ as measured at different red-shifts, differs {\em continuously} from the prediction of the model at those red-shifts, up to the error bars in observations. In fact, there is some (inconclusive) evidence in the current data \cite{holicow, Run, Dainotti, Dainotti1, Dainotti2, Krishnan, Krishnan2} which suggests that the measured $H_0$ might indeed be running with the data redshift $z$. Note that $H_0$ tension was originally noted \cite{Intro0,Intro1,Intro2,Intro3,Intro4,Intro5,Intro6,Intro7,Intro8, Intro9} as a sharp $\sim 5\sigma$ mismatch between $H_0$ measured at low redshifts in our vicinity and that inferred from $z\sim 1100$ via Planck. This result only contrasts the smallest and largest redshifts, but \cite{holicow, Run, Dainotti, Dainotti1, Dainotti2, Krishnan, Krishnan2} raise the possibility of a steadily varying trend in the inferred $H_0$ as one moves from low to high redshift data.  But of course, the data on this is still tentative, so we point this out here merely as an illustration of principle. 


Because our discussion is in many ways quite general, we will initially phrase $H_0$-tension in the general context of models within the FLRW paradigm, restricting to $\Lambda$CDM only for specific discussions eventually (see also \cite{Run}). Under the assumption that the universe is homogeneous and isotropic (and therefore FLRW), we wish to characterize what constitutes a failed FLRW model that results in $H_0$-tension. As outlined already, the way to understand Hubble tension in this setting is as an incompatibility between observations at $z$, and the model for matter-energy as a function of $z$. We will develop this idea in the next section, and find that it immediately leads to some clean and general conclusions. In particular, a decreasing/increasing $H_0$ with redshift is related to an over/underestimate in the {\em total} effective equation of state of the matter\footnote{In this paper, by matter we will mean any perfect fluid, including multi-component fluids.} content used to model the Universe\footnote{In fact, {\em any} late time resolution of Hubble tension within the FLRW paradigm in Einstein gravity, must fall into this general scheme.}. We will express the total Hubble tension between the early and late $z$ measurements in terms of integrals of this over-estimate. In subsequent subsections we consider some simple toy examples to demonstrate these points. To emphasize that $H_0$-tension and its running are more general than $\Lambda$CDM, we illustrate things in the context of a simple dust model. 

Most of our discussions in this paper are within the conventional FLRW paradigm, and therefore  they deal with the running of $H_0$ with red-shift (only). But in an {\em en passant} section, we note the obvious fact that it is easy to generalize this to diagnose deviations away from FLRW. The way to do this is straightforward, one simply has to keep track of data not just as a function of redshift but also as a function of angles in the sky. In {\em any} FLRW model (say $\Lambda$CDM), the data from an underlying anisotropic Universe would lead to an anisotropic $H_0$ in the sky and will serve as a diagnostic for deviations from the model. We mention this, because there are hints in the literature that there may be cracks in the FLRW paradigm via a breakdown of the cosmological principle at late times \cite{Krishnan2, Aniso1, Aniso2, Aniso3, Aniso4, Aniso5,Aniso6, Aniso7,Aniso8,Aniso9,Aniso10}. It is very important to conclusively settle this issue, considering its foundational nature.

In later sections we work more closely with data, and discuss flat $\Lambda$CDM -- but even there, our primary goal will be to illustrate the idea, and {\em not} to do detailed phenomenology of the most up to date datasets\footnote{It is evident that more data is needed to cleanly resolve the various recent tensions in current cosmology. Our present contribution will unfortunately be of no help in this regard.}. In this second half of the paper, our aim is to use the previous philosophy to design $H_0$-diagnostics that are adapted to specific observables. In particular, in writing \eqref{start} and viewing the RHS as a diagnostic as in \cite{Run}, we are using $H(z)$ as the observable. However, many of the natural observables in cosmology are (various kinds of) distances. So it is more useful in many situations to write down $H_0$-diagnostics directly in terms of them. This is straightforward enough to do, and to illustrate this very concretely we will write down $H_0$-diagnostics that are adapted to different BAO observables. We will Gaussian reconstruct $H_0$ with the corresponding BAO data within the flat \LCDM\ model to show how the running works. We will also find that in some cases the diagnostic takes simple forms -- eg., when the radiation can be neglected in \LCDM\, the diagnostic is a simple hypergeometric function, if we are working with comoving distance data. 

To summarize: a suitably defined (running) $H_0$ can be used as a diagnostic that is adapted to any kind of observational data and any specific model within the FLRW paradigm\footnote{It can also (trivially) detect late time anisotropic deviations from the FLRW paradigm.}. The simplicity of the diagnostic for general observables is a consequence of the universality of $H_0$ in FLRW models: it is an integration constant, and not a model parameter. For example, for the complicated observables we consider in this paper in our discussions of BAO, it would be hopeless to try and build a diagnostic based on model parameters.

\section{Hubble Tension and a ``Softer" FLRW Universe}

Our starting point is the observation that the Friedmann equation for general matter can be written as
\bea
H_0^2= H^2 \ a^3 \ \exp\left( -3 \int_a^1 w(a')\frac{da'}{a'}\right),  \label{key}
\eea
where  $w(a)\equiv \frac{p(a)}{\rho(a)}$. This expression is an extremely general statement about the FLRW paradigm: the only assumption we have made here is to take the spatial curvature $k=0$ (we will consider more general $k$ in the next subsection). We have also set $a(t_0)=1$ at the present epoch $t_0$, which can be done without loss of generality. Note that the left hand side is by definition a constant within the FLRW ansatz. Let us also emphasize that we have not made any assumptions about the matter, other than that it is a perfect fluid. In particular, it can have multiple components.

We will consider an ideal scenario for observations, where observational data allows us to measure $H(a)$ at arbitrary values\footnote{Note that within the FLRW ansatz, knowing redshift $z$ fixes $a=\frac{1}{1+z}$ purely kinematically. In other words, $H$ can be viewed as a function of $z$ or $a$.} of $a$. Since $H$ and $a$ can be viewed as observable quantities, it immediately follows from our equation above, that the only way we can have $H_0$ run (assuming our FLRW assumption about the Universe is correct), is if we have made an error in what we believe is our $w(a)$. This is because $a$ (and $H$) should fix the $w(a)$ via Friedmann equations and once it is fixed, (\ref{key}) with a constant left hand side should be an identity. 

With this basic understanding, now we proceed to try and understand how exactly a running $H_0$ tension can arise. 

For clarity going forward, lets distinguish the {\em true} physical  parameters (eg., observed quantities from hypothetical errorless observations) of an FLRW Universe from the parameters of the FLRW {\em model} we will use to describe that Universe. The former will be denoted without subscripts, while we will use $a_{model}, H_{model}$, $p_{model}, \rho_{model} $ and $w_{model}$ to denote the model quantities. Running $H_0$-tension is then the statement that the quantity $h_0$ defined by\footnote{In writing the following equations, we have assumed that the value of $a(t_0)=1$, which enforces that $h_0^2(z=0)=H_0^2$. In principle, we can also consider the possibility that $H_0$ and $h_0(z=0)$ do not match: this will merely shift the $h_0(z)$ curve by a constant amount, equal to the difference.}
\bea
h_0^2(a) := H^2(a) \ a^3 \ \exp\left( -3 \int_a^1 w_{model}(a')\frac{da'}{a'}\right), \label{diag}
\eea
where $w_{model}(a)\equiv \frac{p_{model}(a)}{\rho_{model}(a)}$, has a $z$ (or equivalently $a$) dependence, indicating that the physical observed parameters $H$ and $a$ are not consistent with the matter we have chosen in the model as captured by $w_{model}(a)$. Note the crucial fact that Hubble tension is only possible because we are implicitly comparing two FLRW models (or one model and real-world data). In any given model, $H_0$ is always necessarily a constant.

We can also write the above equation as
\bea
\frac{h_0^2}{H_0^2} = \exp\left( -3 \int_a^1 \Delta w(a')\frac{da'}{a'}\right),
\eea
where $\Delta w(a)\equiv w_{model}(a)-w(a)$. Clearly, now it is no surprise that the left hand side can have non-trivial $z$ dependence: our model is 
wrong. It follows right away that what one calls $H_0$-tension, $\Delta h_0$, is then simply given by
\bea
\Delta h_0 &\equiv & h_0(z_{max})-h_0(z=0) \\
&=& H_0 \ \exp\left( -\frac{3}{2} \int_{a_{max}}^1 \Delta w(a')\frac{da'}{a'}\right)-H_0,
\eea
where $a_{max}=\frac{1}{1+z_{max}}$, and $h_0(z=0) =H_0$ is the value obtained for the present Hubble parameter from (the vanishing limit of) small redshift observations. We have managed to find a simple expression for $H_0$ tension in terms of the discrepancy in the matter content of the model and the real world. 

To illustrate that this is quite a useful equation, lets make an observational prejudice about $h_0$ that may be true in the context of $H_0$-tension in the flat $\Lambda$CDM model (see eg., \cite{holicow, Run, Dainotti, Dainotti1, Dainotti2, Krishnan, Krishnan2}): we will assume that $h_0$ decreases with increasing $z$ or equivalently, that it increases with increasing $a$:
\bea
\frac{d h_0}{d a} > 0.
\eea
It is trivial to check that this translates to the statement that
\bea
\Delta w(a) > 0, \ {\rm or \ equivalently,} \ w_{model}(a) > w(a).
\eea
In other words, a decreasing-with-$z$ trend in the measured value of $H_0$ is precisely equivalent to an over-estimate in the stiffness of the total equation of state. In particular, the conventional $H_0$-tension that is quoted between $z=z_{CMB}$ and nearby supernovae at $z\approx0$ can be directly related to an integrated error in the equation of state, from now to the last scattering surface. One caveat here is that in order to turn angle data to distance data, we need to assume a value for the sound horizon at drag epoch, $r_d$. So this discussion applies only to late time resolutions of the Hubble tension.

\subsection{Including Spatial Curvature}

The inclusion of spatial curvature does not substantively affect the discussions above: we can effectively treat the curvature as another component in the stress tensor. The two independent Friedmann equations can be taken as
\bea
\rho(a)&=&\rho_0 \ \exp \left(3 \int_a^1 \frac{p(a')}{\rho(a')} \frac{da'}{a'} \right), \\
H^2 &=&\frac{\rho}{3}-\frac{k}{a^2},
\eea
where $\rho\left(a(t_0)=1\right) \equiv \rho_0$. Defining 
\bea
\frac{\rho_{crit}}{3} \equiv H_0^2=\frac{\rho_0}{3}-k,
\eea
and
\bea
\Omega_{\rho} \equiv\frac{ \rho_0}{\rho_{crit}}, \ \ \Omega_k=-\frac{k}{\rho_{crit}/3}.
\eea
so that 
\bea
\Omega_0\equiv \Omega_\rho + \Omega_k=1,
\eea
we can get the expression for $H_0$ as
\bea
H_0^2=\frac{H^2}{\frac{\Omega_\rho}{a^3} \ \exp\left(3 \int_a^1 w(a') \frac{da'}{a'}\right)+\frac{\Omega_k}{a^2}}.
\eea
Again, the numerator can be viewed as capturing data 
and the denominator, as capturing model. It is easy to convince oneself that by treating curvature as a fluid of equation state $\omega_k(a) =-1/3$, we can re-interpret this in the same form as before. 

\subsection{Toy Example: $H_0$-tension in a Dust Model}

To drive home the point that $H_0$-tension is a ubiquitous phenomenon when an FLRW universe does not match with the FLRW model that is supposed to describe it, lets consider a simple example. 

Let us assume that the equation of state of a fictitious Universe is slightly softer than dust, but in their infinite simplemindedness, the denizens of said Universe are trying to model it with dust. In other words, the FLRW equations for the ``real'' parameters of the Universe are of the form
\bea
\frac{d \rho}{\rho} = -3 \ (1-\Delta w(a))\ \frac{da}{a}, \ \ 
H^2= \frac{\rho}{3},
\eea 
whereas the model assumes that it is of the form
\bea
\frac{d \rho_{model}}{\rho_{model}} = -3 \ \frac{da_{model}}{a_{model}}, \ \ 
H_{model}^2= \frac{\rho_{model}}{3}.
\eea
Here $\Delta w(a)$ is taken to be positive valued, which means that the ``real" world is softer than the dust model. To write down explicit expressions and make plots, we will take the convenient form
\bea
\Delta w(a) = \epsilon \frac{a}{a(t_0)}, \label{EoS}
\eea
for some small $\epsilon$. This form is taken purely for convenience in doing our integrals, any other choice would do as well\footnote{Note that in order to do the integrals explicitly, it is useful to have this expressed in terms of the scale factor. The knowledge of the equation of state is equivalent to this information -- it basically captures how a particular component of matter redshifts/dilutes with the scale factor.}. For convenience, we will also assume that close to the Big Bang, the error in the model vanishes, this can also be relaxed if one wishes. Going through the calculation, we get the simple result
\bea
h_0^2(z)=H_0^2 \ \exp \left(-\frac{3\ \epsilon \ z}{1+z}\right),
\eea
which leads to the decreasing tendency of $h_0$ with increasing redshift as promised. The corresponding plot of $h_0$ vs $z$ for a representative value of $\epsilon$ (and $H_0$ taken to be 0.07) illustrates the dust-model version of $H_0$-tension.
\begin{figure}[h]\centering
\hspace{-5mm}
\includegraphics[scale=0.45]{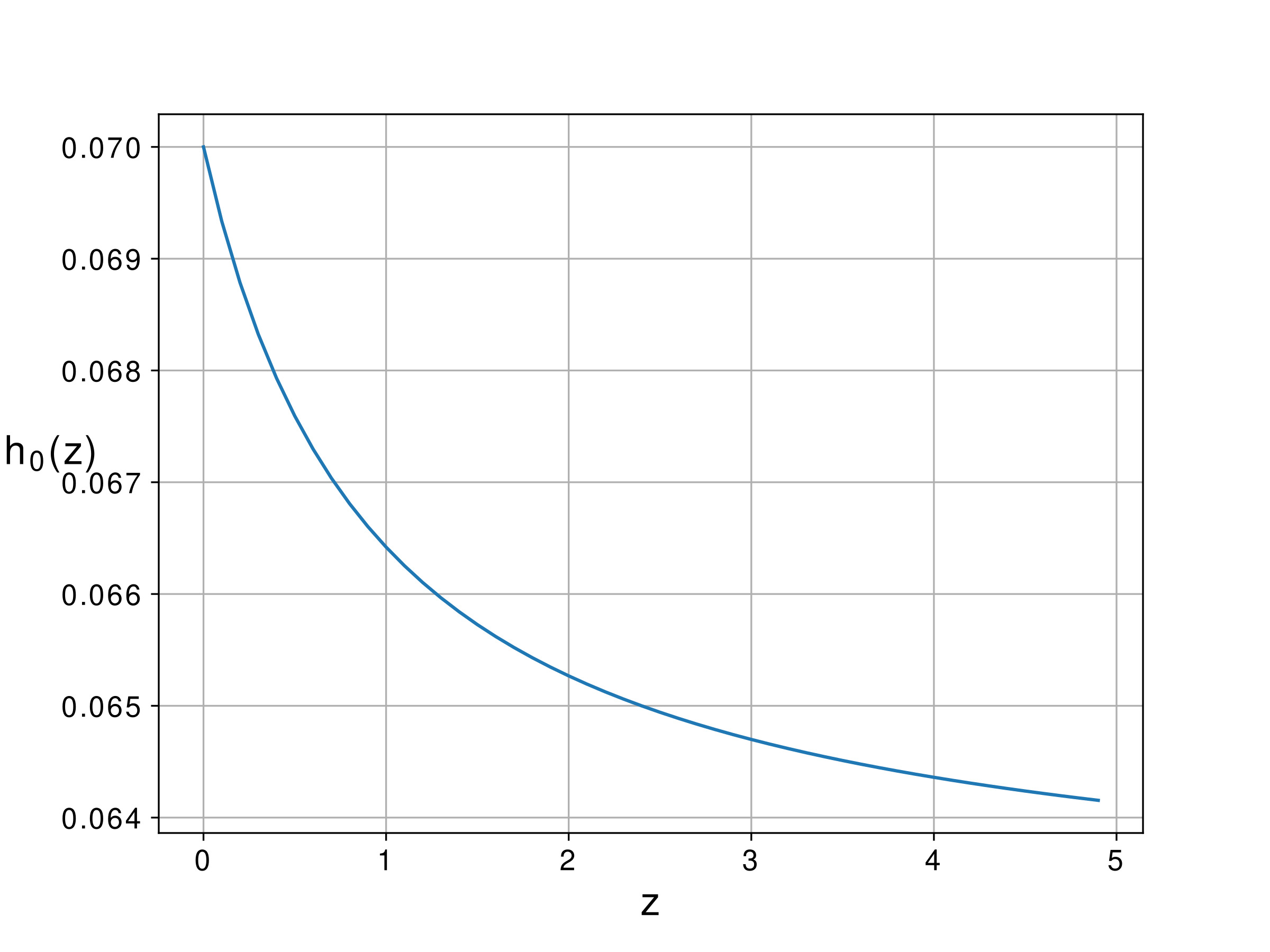} \\
\caption{Plot of $h_0$ vs. $z$ in our dust model.}
\label{XX}
\end{figure}

\section{Aside on Late-Time Anisotropies}

The persistence of tensions between early and late time observations of the Universe has lead to something like a crisis in present day cosmology. One possibility that has been raised in this context is that the deviations from the FLRW paradigm at late times, at redshifts $z \lesssim \mathcal{O}(1)$, may be bigger than previously anticipated. It has been suggested that the flow that our Galaxy is part of, in the direction of the CMB dipole, extends to much higher red-shifts than the $\sim$ 100 MpC that is usually viewed as the coarse-graining scale of the FLRW paradigm\footnote{However, let us point out that this claim suffers from a fine-tuning problem. If the CMB dipole is {\em not} just due to the typical velocity of galaxies in clusters, then it is not clear why the CMB dipole velocity $\sim 600$ km/s is of the same order as that of observed velocities of galaxies in virialized clusters $\sim \mathcal{O}(1000)$ km/s. We will have more to say about this in \cite{KMS}.}. Evidence for this has been accumulating in the literature \cite{Krishnan2, Aniso1, Aniso2, Aniso3, Aniso4, Aniso5,Aniso6, Aniso7,Aniso8,Aniso9,Aniso10}, but the conclusion is still unclear\footnote{See \cite{Soda1,Soda2} for some recent discussions on anisotropic inflation in the {\em early} Universe.}. 

Given this, let us pause here to note that in principle, our diagnostic can also be used to detect such late time anisotropies. All one has to do is to keep track of the angle-dependence of the data on the sky. In place of \eqref{start}, for example, we will have 
\bea H_0(z,\theta,\phi):=\frac{H(z,\theta, \phi)}{\sqrt{\Omega_{m0}(1+z)^3+1-\Omega_{m0}}}. \label{ani-start} 
\eea
where we now are keeping track of the sky location of the $H(z)$ data. If the left hand side is not a constant on the celestial sphere, we can conclude that there are anisotropies. It should be clear that a similar generalization will hold for all the $H_0$-diagnostics adapted to the various observables we will discuss in this paper. Of course, the real bottleneck is data quality and quantity here. 

One small comment perhaps worth making here is that the specific FLRW model one uses in \eqref{ani-start} is not particularly important if one is looking for anisotropies, since {\em any} FLRW model is isotropic by definition. We have written the expression above for $\Lambda$CDM when radiation can be neglected, but this is merely for concreteness. The broader reason for this robustness is that anisotropies rule out the entire FLRW paradigm, and not just one model in the paradigm. 

\section{$H_0$-Diagnostics Adapted to Distance Observables}

Our discussion in the previous section started off with \eqref{key}, and the $H_0$-diagnostic was the one presented in \eqref{diag}. $H$ on the right hand side of \eqref{diag} is a piece of observational data, and the rest of the right hand side is part of the specification of a model. If the left hand side was not a constant, then we would interpret that as the running of $H_0$ and we would conclude that the model was wrong. 

In many cases of interest in cosmology, the direct observables are often distances of various kinds and not $H(z)$ directly. Of course we could take derivatives of distances to obtain $H(z)$ but this would expand the error bars, and therefore we would like to avoid it. Instead, we would like to have $H_0$-diagnostics that are directly adapted to the observational data. In this section we will show that this is easy to do, and we will exhibit it with concrete example diagnostics adapted to distance observables suitable for BAO data. We emphasize that our aim in the following Gaussian reconstruction plots is to provide illustrative examples for the diagnostic -- data quality and true phenomenology are not our focus.



\subsection{Comoving Distance Diagnostic}

\begin{table*}[htp!]
\caption{\label{tab:my_label}{The $D_{M}/r_{d}$ data we use for the plots in Figures 2 and 3.}}
\begin{ruledtabular}
\begin{tabular}{ c c c c }
         
Name & z &  $D_{M}/r_{d}$ &  References  \\ [0.5ex] 
\hline

BOSS & $0.51$ &  $13.36 \pm 0.210$  & S.~Alam {\it et al.} \cite{COMOV} \\

 & $0.57$ &  $14.945 \pm 0.210$ & E.~Aubourg {\it et al.} \cite{BAODATA} \\
\hline
eBOSS & $0.70$ &  $17.86 \pm 0.33$  & S.~Alam {\it et al.} \cite{COMOV} \\

\hline
LyaF & $1.48$ &  $30.69 \pm 0.80$  & S.~Alam {\it et al.} \cite{COMOV} \\

 & $2.33$ &  $37.6 \pm 1.9 $  & S.~Alam {\it et al.} \cite{COMOV} (Ly$\alpha$-Ly$\alpha$)\\

 & $2.34$ &  $36.489 \pm 1.152$ &  E.~Aubourg {\it et al.} \cite{BAODATA} (combined LyaF) \\

 & $2.35$ &  $36.3 \pm 1.8$  & M.~Blomqvist {\it et al.} \cite{Lyaf1}\\
 
& $2.36$ &  $36.288 \pm 1.344$ & E.~Aubourg {\it et al.} \cite{BAODATA}  \\
\end{tabular}
\end{ruledtabular}
\label{Table 1}
\end{table*}

Comoving distance is given by the relation \cite{ Weinberg}:
\bea
D_{C}(z) = c\int_{0}^{z}\frac{dz'}{H(z')}. \label{comovD}
\eea
Comoving distance is a candidate observable in cosmology, and we wish to write down an $H_0$-diagnostic adapted to that. For convenience when making plots, we have changed our independent variable from $a$ to $z$ compared to section II. In this language, we can write the key relation \eqref{key} as
\bea
H(z) = H_{0}\ E(z),
\eea
where
\bea
    E(z) = \text{exp}\Big[\,\frac{3}{2}\int_{0}^{z} dz'\frac{(1+w(z'))}{1+z'} \Big]. \label{model}
\eea
$E(z)$ is the quantity that captures the model. Once the model parameters are specified the integral can be explicitly evaluated. In $\Lambda$CDM for example, at late times when radiation can be neglected it takes the form
\bea
  E^{M}(z) = \sqrt{(1 - \Omega_{m0}) + \Omega_{m0}(1+z)^{3}}. \label{lateLCDM}
\eea 
Running of $H_{0}$ can then be captured by a diagnostic adapted to the comoving distance data via
\bea
H_{0}(z):= \frac{c}{D_{C}(z)}\int_{0}^{z}\frac{dz'}{E (z')}. \label{4}
\eea
where $D_C(z)$ is the data, and $E(z)$ captures the model. The non-constancy of this quantity will serve as the null diagnostic. Slightly more explicitly, we can write
\bea
   H_{0}(z):= \frac{c}{D_{C}(z)}\int_{0}^{z} dz' \text{exp}\Big[-\frac{3}{2}\int_{0}^{z'} dz''\frac{(1+w(z''))}{1+z''} \Big]. \nonumber \\
\eea
Specializing to the late time $\Lambda$CDM model above, this can be explicitly evaluated to be a hypergeometric function:
\begin{widetext}
\begin{eqnarray}  
    H_{0}(z):= \frac{2 \ c}{D_{C}(z)\ \Omega_{m,0}^{5/6}} 
    \Bigg({}_{2}F_{1}\Big[\frac{1}{6},\frac{1}{2},\frac{7}{6},\frac{\Omega_{m,0}-1}{\Omega_{m,0}}\Big]  - \frac{1}{\sqrt{1+z}}\ {}_{2}F_{1}\Big[\frac{1}{6},\frac{1}{2},\frac{7}{6},\frac{\Omega_{m,0}-1}{(1+z)^{3}\Omega_{m,0}}\Big]\Bigg).  \nonumber\\
    \label{8}
\end{eqnarray}
\end{widetext}

So far we have not picked any particular value of $\Omega_{m,0}$. Choosing $\Omega_{m,0}$ completely specifies $\Lambda$CDM at late times. In all the plots below we use the value of $\Omega_{m,0} = 0.303^{+0.018}_{0.016}$ taken from Planck \cite{Planck2018}.

The quantity $D_{C}(z)$ is the line of sight comoving distance \cite{Distance}. What is more directly connected to the BAO data for which we present the plots, is the transverse comoving distance $D_{M}(z)$  \cite{Distance}. When the spatial curvature is zero, as in the flat $\Lambda$CDM model above, the two quantities turn out to be the same $D_{M} = D_{C}$. This is what we have used in the plots. But it should be clear from our discussion in section II that including spatial curvature is entirely straightforward.

\begin{figure}[htp!]
    \centering
    \includegraphics[height=6cm]{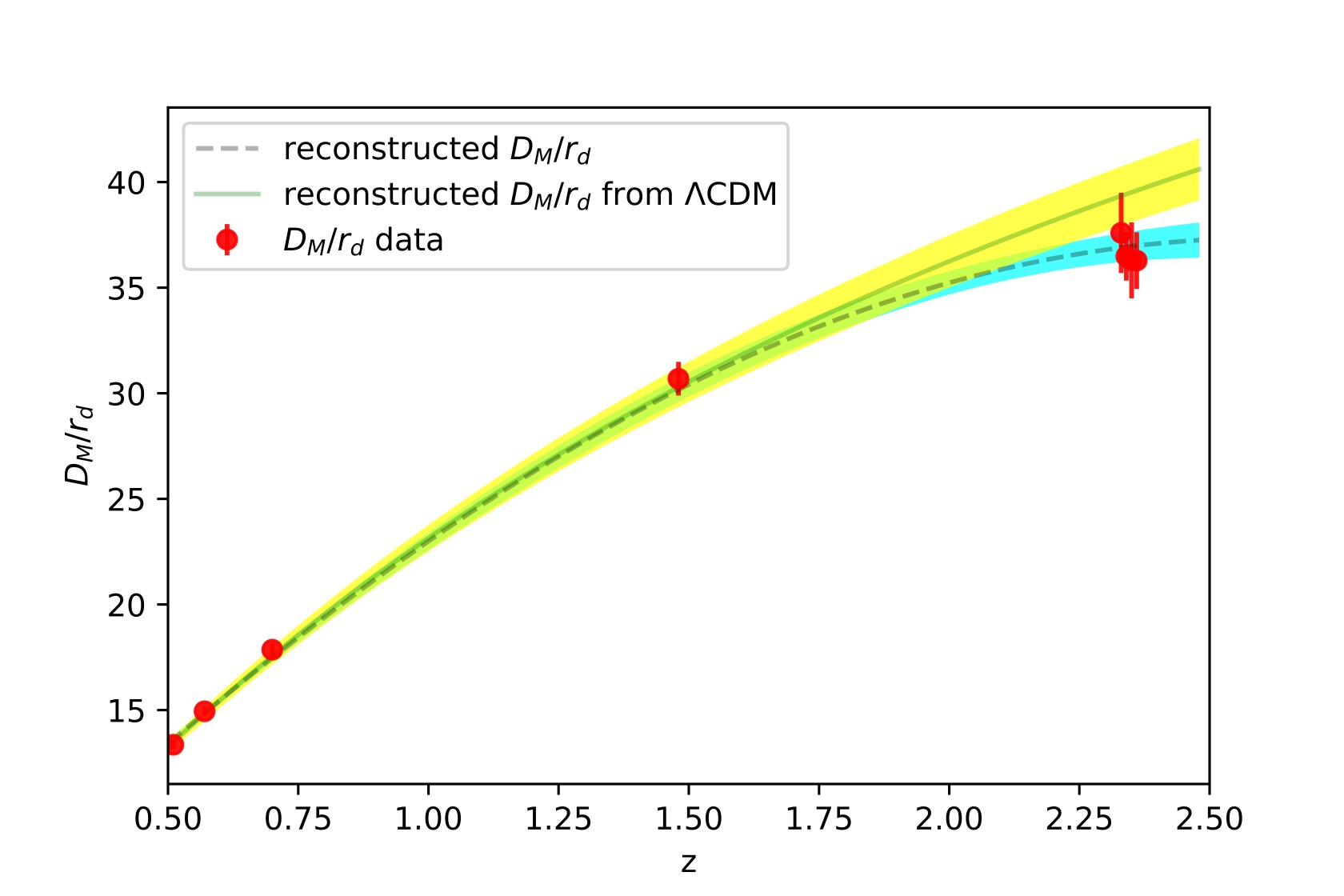}
    \caption{$D_{M}/r_{d}$ reconstructed from the data in TABLE \ref{Table 1}}
    \label{FIG1}
\end{figure}

\begin{figure}[htp!]
    \centering
    \includegraphics[height=6cm]{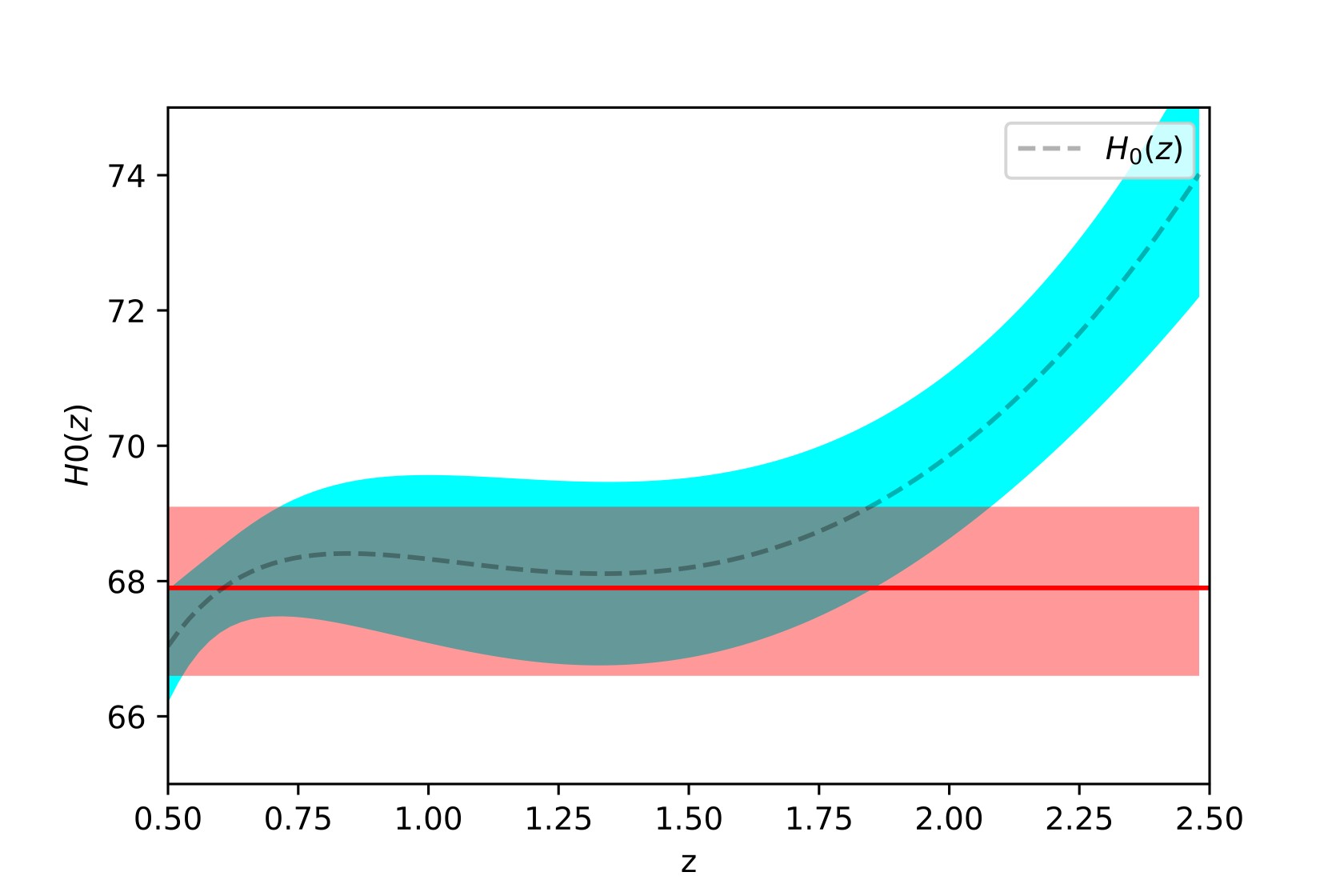}
    \caption{The  inferred  value  of $H_{0}$ from our comoving distance diagnostic  in  units  of km/s/Mpc.  Running of $H_{0}$ from the $D_{M}/r_{d}$ data in TABLE \ref{Table 1}}
    \label{FIG 2}
\end{figure}

To obtain our plots, we need to extract  $D_M (=D_{C})$ from BAO datasets. The BOSS Galaxy surveys and the Lyman $\alpha$ observations in \cite{COMOV,BAODATA} give us $D_{M}/r_{d}$, see  Table \ref{Table 1}. These are basically angle data, and in order to translate them into distances we need to take a fiducial value for the sound horizon at drag epoch, $r_d$. We take this standard ruler length to be $r_{d} = 147.05 \pm 0.3$ MpC \cite{Planck2018} in our $H_0$-diagnostic plot. To make the plots, we first Gaussian reconstruct the $D_M/r_d$ data in the redshift range we are interested in using the handful of data points in Table \ref{Table 1}. This is shown in Figure \ref{FIG1}. With this data and the diagnostic \eqref{8} the $H_0$ running is plotted in Figure \ref{FIG 2}. 

\subsection{Comoving Volume Distance Diagnostic}

As another example of an observable that can be translated into an $H_0$-diagnostic, we consider another BAO angle, 
\bea
1/d_z\equiv D_{V}/r_{d}.
\eea
The quantity $D_{V}$ is called volume comoving distance, and it is defined as
\begin{equation}
     D_{V}(z) = \Big[c\,z\frac{D^{2}(z)}{H(z)}\Big]^{1/3}, \label{eq:2.2}
\end{equation}
where
\begin{equation}
    D(z) \equiv \frac{c}{H_{0}}\int \frac{dz'}{E(z')}.
\end{equation}

\begin{table*}[htp!]
\caption{\label{tab:my_label}{The $d_{z}$ data that we use for the plots in Figures 4 and 5. (In some papers $1/d_z$ is quoted.) }}
\begin{ruledtabular}
\begin{tabular}{ c c c c }
         
Name & z & $d_{z}$  &  References  \\ [0.5ex] 
\hline

6DFGS & 0.106 & $0.336 \pm 0.015$ & C.~Blake {\it et al.} \cite{Wiggle}  \\
\hline
SDSS DR7 
& 0.20 &  $0.1905 \pm 0.0061$ & C.~Blake {\it et al.} \cite{Wiggle} \\

& $0.35$ &  $0.1097 \pm 0.0036$ & C.~Blake {\it et al.} \cite{Wiggle}\\
\hline
WiggleZ & $0.44$ &  $0.0916 \pm 0.0071$ &  C.~Blake {\it et al.} \cite{Wiggle} \\

 & $0.60$ &  $0.0726 \pm 0.0034$ &  C.~Blake {\it et al.} \cite{Wiggle} \\

 & $0.73$ &  $0.0592 \pm 0.0032$ &  C.~Blake {\it et al.} \cite{Wiggle} \\
\hline
MGS & $0.15$ & $0.223 \pm 0.008$ & E.~Aubourg {\it et al.} \cite{BAODATA} \\
\hline
BOSS & $0.32$ &  $0.1181 \pm 0.0023$ & E.~Aubourg {\it et al.} \cite{BAODATA} \\

\hline
eBOSS & $0.85$ &  $0.0546 \pm 0.0018$ & S.~Alam {\it et al.} \cite{COMOV} \\
\end{tabular}
\end{ruledtabular}
\label{Table 2}
\end{table*}

\begin{figure}[H]
    \centering
    \includegraphics[height=6cm]{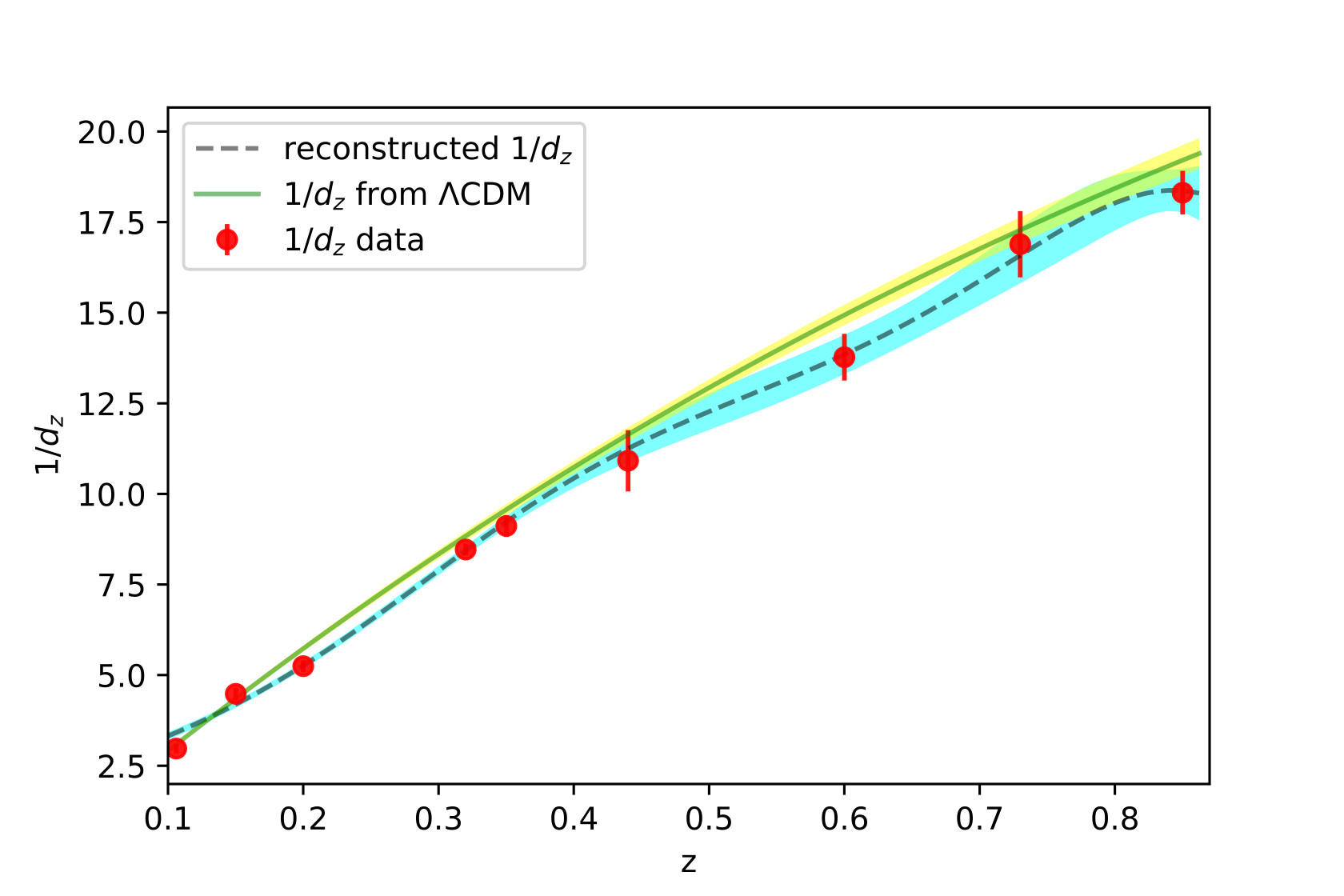}
    \caption{Reconstructed $d_{z}$ from the data in Table \ref{Table 2}}
    \label{fig:my_label1}
\end{figure}
\vspace{-0.3in}
\begin{figure}[H]
    \centering
    \includegraphics[height=6cm]{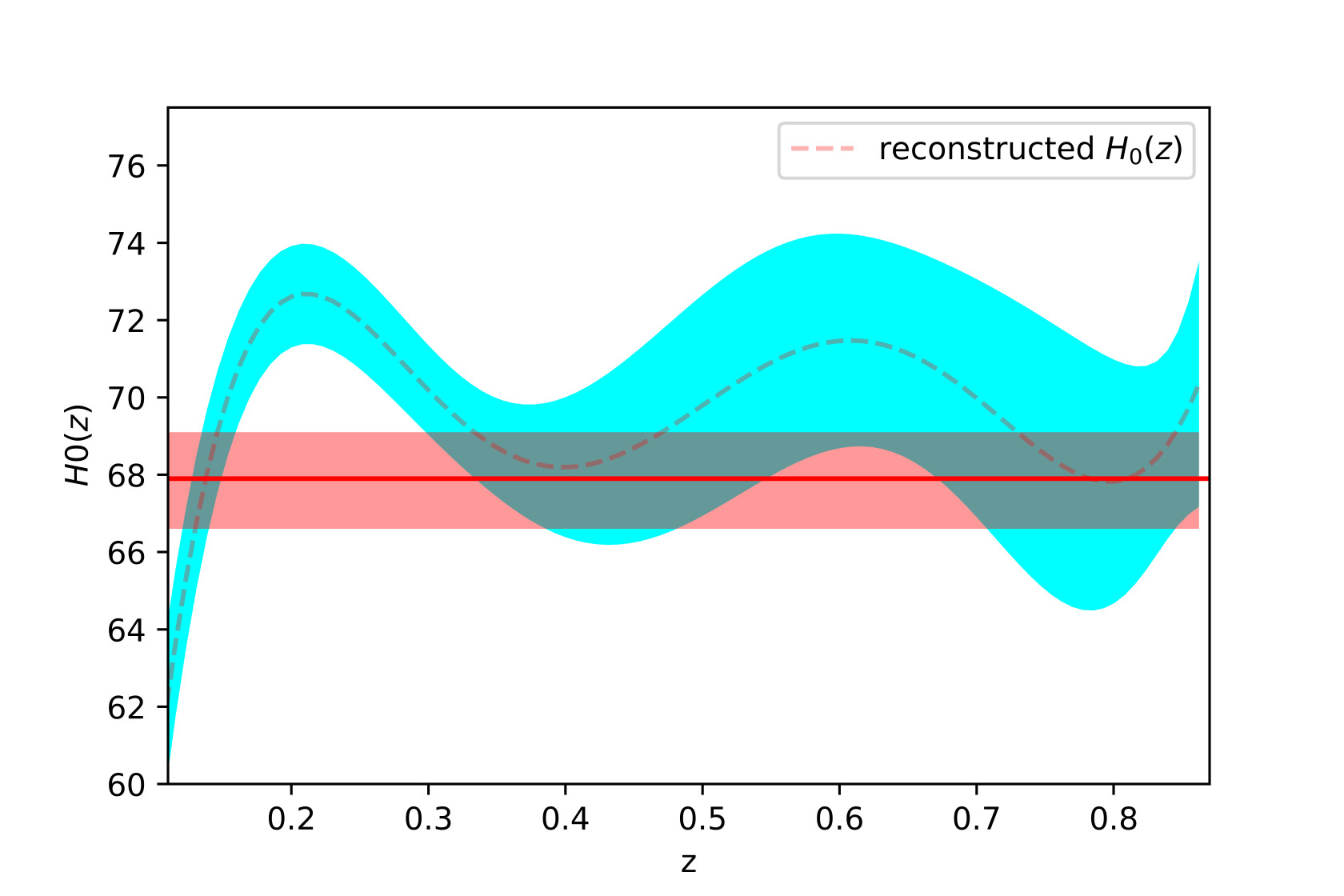}
    \caption{The  inferred  value  of $H_{0}$ from our comoving volume distance diagnostic  in  units  of km/s/Mpc. Running of $H_{0}$ from the $d_{z}$ data in Table \ref{Table 2}}
    \label{fig:my_label2}
\end{figure}

As we discussed in the previous subsection, the expansion history  $H(z) = H_{0} E(z)$ captures the model information via $E(z)$. Hence, $H_{0}$ from the Eq. \eqref{eq:2.2} leads us to a diagnostic
\bea
    H_{0}(z):= \frac{c}{D_{V}(z)}\Big[\frac{z}{E(z)}\Big(\int \frac{dz'}{E(z')}\Big)^{2}\Big]^{1/3}. \label{16}
\eea
Despite the somewhat complicated look, let us emphasize that for a given model \eqref{model}, the right hand side is an {\em explicit} function of the model parameters (as well as the volume distance dataset). Check of running $H_0$ proceeds as before, we do this for $\Lambda$CDM at late times using \eqref{lateLCDM}. 

To make the plots, we use the $d_z$ data presented in Table \ref{Table 2}. Again we use Gaussian reconstruction to obtain values of $d_z$ in the desired redshift range, see Figure \ref{fig:my_label1}. With that and using the fiducial $r_d$ as before, this leads us to a comoving volume distance dataset. For late time $\Lambda$CDM \eqref{lateLCDM}, we can then plot the diagnostic using \eqref{16}, this is Figure \ref{fig:my_label2}. 

It should be clear from these discussions that despite the complexity of the observables, it is straightforward to build simple diagnostics based on $H_0$. As mentioned in the introduction, this is a consequence of the fact that $H_0$ is a universal parameter -- an integration constant --  in FLRW models. A diagnostic built from model parameters would not be able to retain this simplicity and universality. 


\section{Outlook}

Observations at two different red-shifts, even when the systematics are under control, can lead to conflict between the corresponding inferred values of  $H_0$. Indeed, this can happen even in a Universe that is exactly FLRW, and need not be a result of local inhomogeneities or anisotropies. This fact can be confusing, because basic cosmology tells us that $H_0$ is tautologically a constant in an FLRW Universe. In this paper and \cite{Run}, we systematically explored what it means for Hubble tension to exist in a general FLRW universe. The tension exists only because there is a (sometimes implicit) comparison between observational data and an underlying model, or two different models. In other words, while model-independent measurements of $H_0$ at low red-shifts are possible, there is no such thing as a model-independent notion of Hubble {\em tension}. 

Reversing the logic of this, we argued that $H_0$ can be used as a litmus test on the observational validity of {\em any} FLRW model. Given cosmological distance data from different red-shifts, we wrote down null hypothesis diagnostics that indicate whether a given model has Hubble tension according to that data. Implicit in it is the observation that if there is tension, it must  be a quasi-continuous function of the data red-shift: $H_0$ tension {\em must} run. In simple models, we presented explicit formulas for our diagnostic, eg., for flat \LCDM\ when radiation contribution can be neglected, the comoving distance diagnostic reduces to a hypergeometric function. This improves on the related recent proposal in \cite{Run} in that it works directly with distance data\footnote{See also \cite{Om, Om1} for the ``Om-diagnostic" that deals with vacuum energy (via the $\Omega_{m,0}$ parameter of $\Lambda$CDM), and not $H_0$.}. If we trust the Planck value for the sound horizon at drag epoch $r_d$, we can write down an explicit expression for Hubble tension in terms of integrals involving the \LCDM\ matter content over red-shifts from now to recombination. This measures an integrated error in the \LCDM\ effective equation of state since last scattering, from which we can extract the slogan: {\em Hubble tension implies a softer universe}. 

$H_0$ has historically been a difficult quantity to measure. Despite this, the recent crisis in cosmology can very directly be traced to our increased ability to measure $H_0$ more reliably\footnote{Let us note however that the existence of the tension has been challenged in \cite{Rameez}: the sky variation of $H_0$ is argued to be already $\sim 10\%$.}. As the quality and quantity of observational data improves over the years, the non-constancy of $H_0$ could be useful as a broadly applicable test of the correctness of cosmological models, irrespective of whether the presently observed Hubble tension is a real feature or not. In other words, the viability of $H_0$ as a  diagnostic is a direct measure of precision, in ``precision cosmology".

Let us conclude by making one observation: it is noteworthy that the current tensions are all reliant on late time data. Interestingly, so is the evidence for the accelerated expansion of the Universe and the cosmological constant. Realizing cosmic acceleration in UV complete theories has turned out to be a thorny problem: recently this has been emphasized in the dS swampland program -- see \cite{Hertzberg, Vafa1} for the initial proposal, \cite{Garg, Vafa2} for the potentially viable version, and \cite{Andriot, Garg2} for further refinements. In light of this, it is conceivable that resolving the tension in late time cosmology may have far-reaching implications.

\section{Acknowledgments}


We thank Roya Mohayaee, Eoin O'Colgain, Ruchika, Subir Sarkar, Anjan Sen, Shahin Sheikh-Jabbari and Lu Yin for related discussions.

\end{document}